\documentclass[11pt]{article}
\usepackage{graphicx}

\def\be{\begin{equation}}
\def\ee{\end{equation}}
\def\ba{\begin{eqnarray}}
\def\ea{\end{eqnarray}}

\begin{document}
\begin{titlepage}
\title{\begin{flushright}\begin{small}    
\end{small} \end{flushright} \vspace{2cm}
The analog of the Hawking effect in BECs\footnote{Plenary talk at ERE2014, Valencia, September 2014}}

\author {Alessandro Fabbri
\thanks{Email:afabbri@ific.uv.es}\\ \small{Centro Studi e
Ricerche Enrico Fermiâ Piazza del Viminale 1, 00184 Roma, Italy}
\\ \small{Dipartimento di Fisica dell'Universit\`a di Bologna,
Via Irnerio 46, 40126 Bologna, Italy}\\ \small{Dep. de F\'isica
Te\'orica and IFIC, Universidad de Valencia-CSIC,}\\ \small{C. Dr.
Moliner 50, 46100 Burjassot, Spain}}

\maketitle
\begin{abstract}
The observation of the Hawking effect from black holes in the astrophysical context is unlikely. However, the analog of this effect is present in condensed matter systems.
We focus on Bose-Einstein condensates, and on a proposal to detect it through correlation measurements.
\end{abstract}
\end{titlepage}\setcounter{page}{2}
\section{Introduction}
\bigskip 

Gravity is by far the weakest interaction and its presence at the microscopic level is usually neglected. Quantum gravitational effects are extremely small, as typically one should go to the Planck scale to measure them. This makes the construction of a quantum theory of gravity a very difficult task and, mainly, a challenge for theorists.

Despite these facts, however, gravity is universal and always attractive. It is  the dominant force in the macroscopic world and determines the large scale structure of the Universe.
Moreover, when matter collapses inside its Schwarzschild radius $r_G=\frac{2GM}{c^2}$ it can form black holes, in which the gravitational field is so strong that light is trapped and cannot escape. They form in the gravitational collapse of massive stars with mass $M\geq 2-3M_{Sun}$, their outer boundary is the event horizon.

Black holes are important for many reasons, in particular because they allow unification of gravity with thermodynamics.
In the early seventies Bardeen, Carter and Hawking \cite{bch} pointed out a surprising analogy between the laws of black hole mechanics
\begin{equation}  
c^2\delta M=\frac{c^2}{8\pi G}\kappa\delta A_H+\Omega_H\delta J\ , \ \ \delta A_H>0\ , .. \end{equation}
(where $M$, $J$ are the black hole's mass and angular momentum and $A_H,\ \Omega_H,\ \kappa$  the horizon's area, angular velocity and surface gravity)
and the laws of thermodynamics. This fact, in particular the area law \cite{haw} $\delta A_H>0$ expressing that the surface of a black hole's event horizon never decreases, led Bekenstein \cite{bek} to propose that $A_H$ gives a measure of the entropy of the black hole, i.e. $S_{BH}\propto A_H$. In order for this analogy to be taken seriously, however, a way was needed to assign a temperature to the black hole. At the classical level this is impossible since a black hole is a perfect absorber and has, therefore, $T=0$. 

This problem was solved by Hawking \cite{hawk}. He used the formalism of quantum field theory in curved space, initially developed by Parker \cite{park}, to show that black holes quantum mechanically emit a thermal flux of particles at the characteristic temperature
\begin{equation}\label{hr}
T_H=\frac{\hbar\kappa}{2\pi ck_B}\ ,
\end{equation}
with $\kappa$  the horizon's surface gravity. He also clarified the basic mechanism responsible for the emission. Virtual pairs are continuously created from the vacuum but annihilate almost instantaneously (after a time of the order of Planck time) leaving, in ordinary situations, no visible effect. Close to the event horizon, however, one member of the pair may get trapped inside the hole, leaving its partner free to propagate away where it is detected. Such particles, called the Hawking quanta, constitute the particles emitted by the black hole. 

Eq. (\ref{hr}) is a remarkable prediction of a quantum effect in gravity. Unfortunately, in realistic situations in which black holes form from gravitational collapse
\begin{equation}\label{hrast} T_H \sim \frac{M_{Sun}}{M} \ K \ll T_{CMB} \sim 3\ K \ ,
\end{equation}
i.e. the Universe' cosmic microwave background temperature is far too high making the experimental detection of Hawking radiation from astrophysical black holes very unlikely. Alternative possibilities are based on the fact that mini black holes (being $T_H \propto \frac{1}{M}$ this effect becomes more important for small masses) may have been produced in the early Universe by quantum effects or at particle accelerators due to the existence of (large) extra dimensions.

\bigskip
\setcounter{equation}{0}\section{Analog gravity}
\bigskip

An important property of Hawking radiation, uncovered by Unruh \cite{unr}, is that it is not specific to gravity. Indeed, this effect has a kinematical origin. Hawking
did not use Einstein's field equations to derive it, his analysis only relies on the details of wave propagation around the black hole's horizon.

A stationary fluid undergoing supersonic motion traps sound waves in its supersonic region in the same way a gravitational black hole traps light. Such configurations, called acoustic black holes, possess an acoustic horizon, which is the surface where the fluid's velocity reaches the speed of sound. 


There is a precise mathematical equivalence between light propagation around the event horizon of a gravitational black hole and sound propagation around the acoustic horizon of an acoustic black hole \cite{blv}. Unruh used it to predict that acoustic black holes will emit a thermal flux of phonons (analog Hawking radiation) from their acoustic horizon at the temperature
\begin{equation}
T_H=\frac{1}{4\pi k_Bc_s}\frac{d(c_s^2-\vec v_0^2)}{dn}|_{hor}\ ,
\end{equation}
where $c_s$ is the sound speed, $\vec v_0$ the fluid's velocity and $n$ is the normal to the horizon. 

This application has various advantages. First of all, the above considerations are valid not only for fluids, but also for other condensed matter systems in the hydrodynamic (long wavelength) approximation. 

From a theoretical viewpoint it is important because it allows to address one weak point in Hawking's derivation, the transplanckian problem \cite{jac}. Due to the huge redshift experienced by the field modes describing the asymptotic Hawking quanta in the propagation between the horizon and infinity, their initial (horizon's) frequency  must have been enormous (i.e. transplanckian). In such a regime, we can no longer trust the formalism of QFT in curved space used by Hawking, we need a quantum gravity calculation. This difficulty makes Hawking's analysis dubious, and  the results uncertain. In this context we have an analog problem (i.e. Hawking radiation relies on frequencies higher than the cutoff scale set by the validity of the hydrodynamical fluid description), however in many cases
a microscopic description 
is available.

The other advantage of this application is to open up new possibilities to test experimentally the analog of the Hawking effect. Two classes of experiments have been realised, water tanks experiments 
\cite{wein} and quantum optics experiments using ultrashort laser pulse filaments \cite{bel}. The interpretation of the  results of these interesting experiments is, however, controversial \cite{mpa, sun}.

\bigskip
\setcounter{equation}{0}\section{Acoustic black holes in BECs}
\bigskip

Bose-Einstein condensates (BECs) are dilute gas of bosons cooled down at extremely low temperatures (of the order of $100\ nK$) and characterized by having (almost) all its  constituents in the same quantum state. In such a state of matter macroscopic quantum phenomena become apparent. 
Predicted in 1924 by Bose and Einstein, they were discovered in 1995 by the groups of Cornell, Wieman and Ketterle using, respectively, atomic gases of rubidium and sodium. 

BECs offer particularly favourable experimental conditions for the detection of the analog of the Hawking effect.  
Indeed on can reach situations in which the Hawking temperature of the BEC acoustic black hole is only one order of magnitude lower than that of the condensate
\begin{equation} T_H\sim 10\ nK \leq T_C \sim 100\ nK \ . \end{equation}
This is a very significative improvement with respect to astrophysics (\ref{hrast}), but it is
not enough to attempt a direct detection of the thermal Hawking's emitted phonons.

Another important advantage of having an analog black hole in the lab is that unlike in the gravitational case, where measurements are necessarily restricted to the exterior of the black hole, no such restriction exists for the interior of an acoustic black hole (which is just a region of supersonic flow). Thus, measurements can be made in exterior as well as in the interior. 
This makes it possible to experimentally detect the analog of the Hawking effect via the pair creation phenomena that uniquely characterises it, i.e. Hawing quanta (in the exterior) - partner (in the interior). 

Using QFT in curved space techniques, valid in the (quantum) hydrodynamic approximation of BEC theory, it was shown \cite{bal} that the correlator of the density fluctuations 
$\langle \delta \hat n(x)\delta \hat n(x')\rangle$
around an acoustic black hole in a BEC displays a characteristic peak for points, respectively, outside and inside the acoustic horizon and which clearly identifies the Hawking effect. This motivated an `ab initio' numerical analysis within the microscopic BEC theory \cite{car} whose results, displayed in Fig. 1, nicely confirmed the hydrodynamical prediction.

\begin{figure}[h]
\centering \includegraphics[angle=90, height=3.8in] {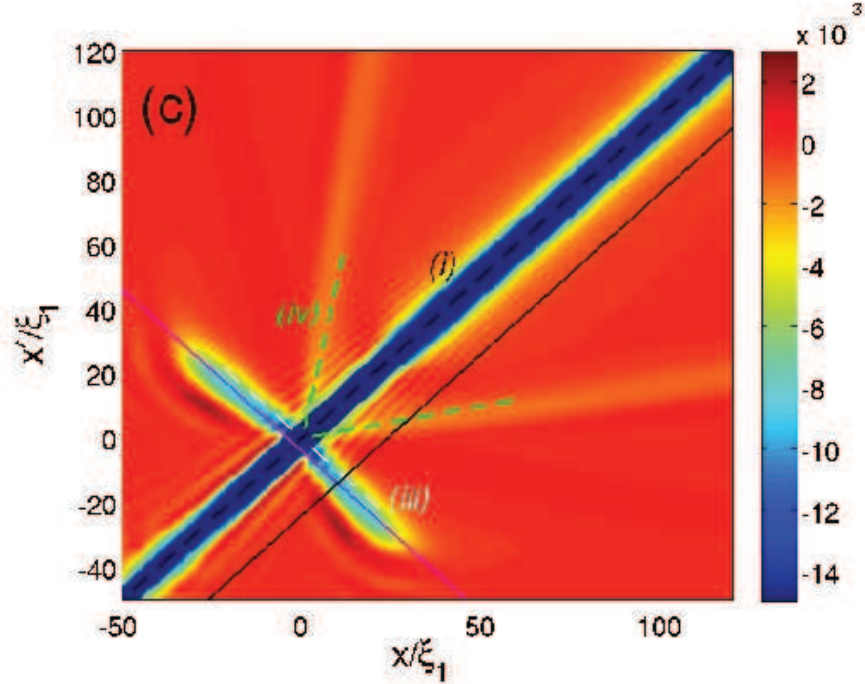}
\caption{Density-density correlator in a 1D acoustic black hole. $x=0$ is the acoustic horizon. Feature iii), (roughly) perpendicular to the main diagonal identifies the Hawking quanta - partner correlations for points on both sides of the acoustic horizon ($xx'<0$).}
\end{figure}

The existence of Hawking radiation in BECs vs. the transplanckian problem is therefore established. Another interesting feature of the numerical results is that 
if we take the fluctuations to be initially in a thermal state (instead of the vacuum) at a temperature bigger than $T_H$ the Hawking signal is not masked by thermal effects, a feature making
the peak in the correlator the `smoking gun' to experimentally identify the analog Hawking radiation.

In addition to black holes, in this context one can reproduce configurations which in gravity are problematic. It is the case of white holes, time reversal of black holes. Whereas black holes are perfect absorbers, white holes expel any matter inside them and do not allow anything to enter from the outside.  The problem in gravity is that they require an initial singularity, and this makes physical predictions about them impossible. 

The analog Hawking radiation requires the existence of negative energy states, and these exist in the supersonic region of both black holes and white holes \cite{balbi}. The big difference between the two cases is that while in black holes the Hawking effect is a low energy effect (it does not depend on the details of the microscopic physics), in white holes it is a high-energy effect and, thus, its signal is system dependent. 

In BECs, the Hawking effect in white holes is crucially 
 determined by the short-scale modification of the linear  (phononic) dispersion relation of the Bogoliubov fluctuations, which
becomes supersonic at high momenta and develops, inside the white hole, a nontrivial (outgoing) zero mode  \cite{may}. 
This happens for  $k\geq  \frac{1}{\xi}$, where $\xi$ is the healing length, the natural length scale of the condensate and the analog of the Planck length in gravity. 
We see in the density correlator of Fig. 2 that in the supersonic region $x,x'<0$ a checkerboard pattern emerges, characterized by a series of oscillations with peaks parallel and antiparallel to the main diagonal. The overall amplitude of these oscillations grows with time, logarithmically if the initial state is the vacuum and linearly in the case of an initial thermal state. It describes the emission of (an infinite number of) zero-frequency waves, and signals a (weak) instability of the system.

\begin{figure}[h]
\centering \includegraphics[angle=90, height=3.5in] {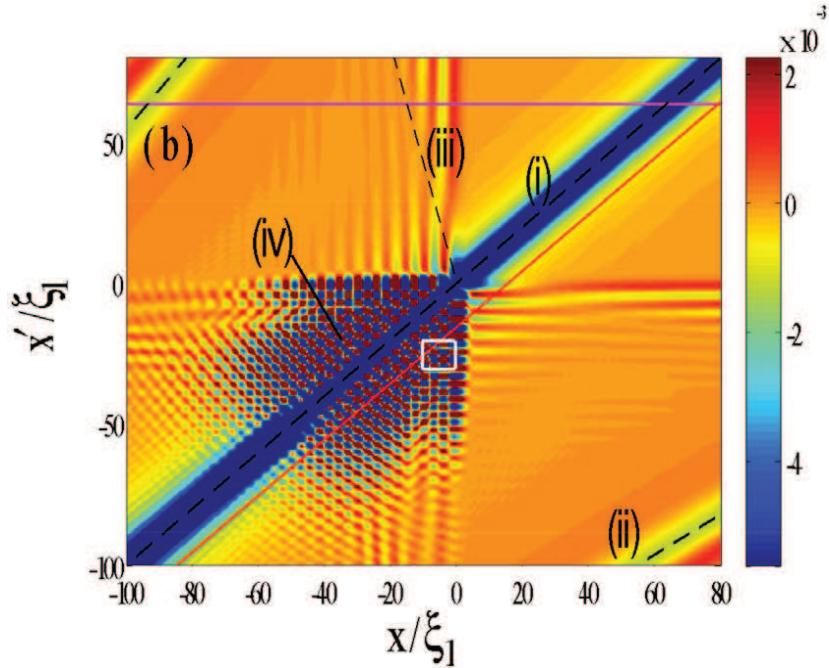}
\caption{The checkerboard patter in the density correlator of an acoustic white hole in a BEC. The oscillations taking place in supersonic region ($x,\ x'<0$) have an overall amplitude that grows with time, signalling a (weak) instability.}
\end{figure}

A dramatic manifestation of this instability takes place in a composite system, made of a black hole and a white hole horizon, that has been realised experimentally in \cite{lah}. In these systems, pairs are created at the black hole horizon, the Hawking quanta in the exterior and its partner in the interior. The partner propagates towards the white hole horizon where, due again to the supersonic character of Bogoliubov dispersion, it `bounces' and comes back to the black hole horizon generating another Hawking quanta. This process repeats itself again and again and results in a laser-type instability \cite{fin}. This makes the system dynamically unstable and the emitted Hawking flux exponentially amplified. Evidence of this effect has been observed recently in \cite{jeff}.

\bigskip
\setcounter{equation}{0}\section{Dynamical Casimir effect and momentum correlators}
\bigskip

The numerical results for the density correlator in acoustic black holes in \cite{car} displayed, at early times, a transient term, the feature ii) in Fig. 3. This additional peak  does not depend on the inhomogeneities and the presence of the horizon. It is there because of time dependence, and appears because in the simulations the acoustic black hole was dynamically created. This feature is a manifestation a distinct pair creation taking place even in homogeneous systems, the  dynamical Casimir effect \cite{caru}. 

\begin{figure}[h]
\centering \includegraphics[angle=0, height=3.8in] {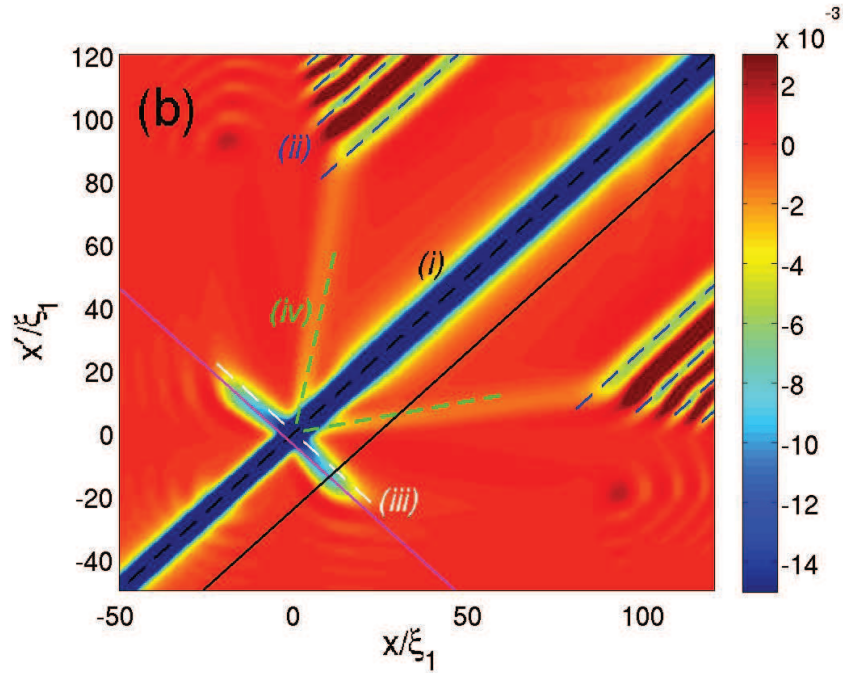}
\caption{The early time simulations in acoustic black holes showed the existence of a transient term, the feature ii), which disappears at late time, Fig. 1, and is due to the dynamical
Casimir effect.}
\end{figure}

An experiment was carried out to measure this effect in \cite{jas}. In it the confining potential of a homogeneous condensate was varied in time (both with a sudden compression and a periodic modulation). This created excitations around the condensate, which were measured with the time of flight technique. 
From the experimental data the correlation function with respect to the atoms' velocities 
was constructed. This displayed a peak along the anti-diagonal, $v=-v'$,  which corresponds to the creation of correlated excitations at opposite momenta $k=-k'$ as predicted by the dynamical Casimir effect. It is worth mentioning that the same effect is responsible for particle creation in the early Universe. More specifically, it would be interesting to mimic it by reproducing in the lab the same de Sitter exponential expansion the Universe experienced during the (early) inflationary era. 

The next step is to reproduce experimentally an acoustic black hole with only one horizon, using the `waterfall' potential of \cite{lar} in which the the horizon and the supersonic region are created by an external step-like potential.
At the theoretical level, we need to adapt the original proposal to study the features of the Hawking effect in the correlator of the fluctuations' occupation number operator
$\langle \delta \hat n(k) \delta \hat n(k')\rangle$ in momentum space 
\cite{boi}.

\bigskip
\setcounter{equation}{0}\section{Infrared properties of the analog Hawking radiation}
\bigskip

A black hole does not emit as a perfect black-body. Indeed, the modes of the Hawking quanta, from the the horizon, reach the far away detector with a probability $\Gamma(\omega)$ that depends on their frequency. Such quantity, called the gray-body factor, enters Hawking's formula for the particles emitted by the black hole and modulates the Planck distribution
\begin{equation}
N_{\omega}=\frac{\Gamma(\omega)}{e^{\frac{\hbar\omega}{k_BT_H}}-1}\ .
\end{equation}

Of particular interest is the low frequency emission.
In gravity, for a black hole in asymptotically flat space $\Gamma$ vanishes at small $\omega$ as $\Gamma\sim A_H\omega^2$, where $A_H$ is the horizon's area \cite{pag}. This allows to eliminate the infrared divergence ($\frac{1}{\omega}$) in the Planck distribution and, moreover, indicates that the emission of low energy particles is suppressed. 

Surpsingly, for a 1D acoustic black hole with just one horizon the graybody factor $\Gamma$ goes to a nonzero constant at low frequency \cite{and}. This behaviour, which is similar to that of a black hole immersed in an expanding de Sitter space, shows that the analog Hawking radiation is dominated by an infinite number ($\frac{1}{\omega}$) of soft phonons. 
One might worry that this infrared divergence will show up also in the density correlator, complicating its use to identify the Hawking effect.
It is however not so, as the density-density correlator stays infrared finite. 
\bigskip
\setcounter{equation}{0}\section{Future directions}
\bigskip

In gravity  it is important to understand how the Hawking effect modifies the classical black hole geometry. The semiclassical Einstein equations, or backreaction equations, are useful to describe the details of the evaporation \cite{fab} at most until the black hole reaches the Planck mass, after which we enter in the realm of quantum gravity. By extrapolating his semiclassical results, Hawking predicted that black holes will disappear and, with them, the information of the star from which they were formed \cite{hawki}. If so, it will have dramatic consequences, as it would imply that from a full quantum gravity perspective the whole process formation and evaporation of a black hole will be described by an S-matrix which is not unitary, in clear contradiction with the principles of quantum mechanics.

It is generally believed that unitarity will be preserved, however the difficult part is to understand how. Recently, it has been conjectured that under a set or reasonable assumptions unitarity of the emitted Hawking radiation will imply that at some (Page) time a high-energy barrier (a `firewall' \cite{alm}) will form at the horizon destroying the Hawking quanta-partner correlations. These qualitative issues can be addressed concretely in BECs, where Hawking quanta-partner correlations can be measured and backreaction equations can be written down which are valid at all scales. 


\bigskip

\end{document}